\documentclass[10pt, conference]{IEEEtran}
\IEEEoverridecommandlockouts
\usepackage{stfloats}
\usepackage{cite}
\usepackage{amsmath,amssymb,amsfonts}
\usepackage{graphicx}
\usepackage{textcomp}
\usepackage{caption}
\usepackage{xcolor}
\def\BibTeX{{\rm B\kern-.05em{\sc i\kern-.025em b}\kern-.08em
    T\kern-.1667em\lower.7ex\hbox{E}\kern-.125emX}}
\usepackage{algorithm}  
\usepackage{algorithmic}
\usepackage{amsmath}
\usepackage{subfigure}
\usepackage{enumerate}
\usepackage{comment}
\usepackage{multirow}
\usepackage{threeparttable}

\begin{document}

\title{Semantic Communication-Empowered \\Physical-layer Network Coding
\vspace{-0.1in}
}

\author{
    \IEEEauthorblockN{Shuai Yang\IEEEauthorrefmark{1}, Haoyuan Pan\IEEEauthorrefmark{1}, Tse-Tin Chan\IEEEauthorrefmark{2}, Zhaorui Wang\IEEEauthorrefmark{3}}
    \IEEEauthorblockA{\IEEEauthorrefmark{1} College of Computer Science and Software Engineering, Shenzhen University, Shenzhen, China}
    \IEEEauthorblockA{\IEEEauthorrefmark{2} Department of Mathematics and Information Technology, The Education University of Hong Kong, Hong Kong SAR, China}
    \IEEEauthorblockA{\IEEEauthorrefmark{3} School of Science and Engineering,  Future Network of Intelligence Institute, \\The Chinese University of Hong Kong (Shenzhen), Shenzhen, China}    
    \IEEEauthorblockA{E-mails: 2110276206@email.szu.edu.cn, hypan@szu.edu.cn, tsetinchan@eduhk.hk, wangzhaorui@cuhk.edu.cn}
    \vspace{-0.4in}
}

\maketitle

\begin{abstract}
In a two-way relay channel (TWRC), physical-layer network coding (PNC) doubles the system throughput by turning superimposed signals transmitted simultaneously by different end nodes into useful network-coded information (known as PNC decoding). Prior works indicated that the PNC decoding performance is affected by the relative phase offset between the received signals from different nodes. In particular, some ``bad'' relative phase offsets could lead to huge performance degradation. Previous solutions to mitigate the relative phase offset effect were limited to the conventional bit-oriented communication paradigm, aiming at delivering a given information stream as quickly and reliably as possible. In contrast, this paper puts forth the first semantic communication-empowered PNC-enabled TWRC to address the relative phase offset issue, referred to as SC-PNC. Despite the bad relative phase offsets, SC-PNC directly extracts the semantic meaning of transmitted messages rather than ensuring accurate bit stream transmission. We jointly design deep neural network (DNN)-based transceivers at the end nodes and propose a semantic PNC decoder at the relay. Taking image delivery as an example, experimental results show that the SC-PNC TWRC achieves high and stable reconstruction quality for images under different channel conditions and relative phase offsets, compared with the conventional bit-oriented counterparts.
\end{abstract}

\vspace{-0.05in}
\section{Introduction}
\label{section1}
Over the past years, physical-layer network coding (PNC) has received significant research attention and has been widely studied in various communication scenarios \cite{zhang2006hot}. As a key technique to overcome wireless interference by mapping superimposed electromagnetic waves into network-coded information, PNC can effectively increase spectrum efficiency and enhance the throughput of wireless networks \cite{zhang2006hot,liew2015primer}. 

The idea of PNC can be most easily illustrated by a two-way relay channel (TWRC) \cite{zhang2006hot}. In a TWRC, suppose that two end nodes, nodes A and B, want to send packets to each other over a wireless medium. Due to the long distance between nodes A and B or their low transmit power, there is no direct signal path between them. They need to communicate with the assistance of a relay node R. Compared with traditional store-and-forward relaying, PNC reduces the total number of time slots to two for exchanging two packets \cite{liew2015primer}. Speciﬁcally, in the first time slot, the two nodes simultaneously send packets to the relay in the same frequency band. In the second time slot, the relay performs PNC decoding on the superimposed received signals and broadcasts back a network-coded packet (a PNC packet) to the nodes. Upon receiving the PNC packet, the two nodes subtract their self-packet from the PNC packet to obtain the packet from another node. As a result, PNC halves the transmission time and doubles the throughput of a TWRC.

In a PNC-enabled TWRC, the successful PNC decoding at the relay is paramount to the overall system performance. Prior works revealed that the PNC decoding performance is affected by the relative phase offset between the signals of simultaneously transmitted packets from the two nodes \cite{pan2017network}. For example, different relative phase offsets lead to different bit error rate (BER) performances in PNC decoding. As will be further elaborated in Section \ref{3.2}, under ``bad'' relative phase offsets, when mapping the superimposed signals to network-coded symbols in PNC decoding, referred to as \emph{PNC mapping}, some constellation points mapped to different network-coded symbols are inevitably overlapped with each other. This leads to symbol ambiguity in the symbol detection process and thus degrades the BER performance.  

Several approaches have been proposed to overcome the relative phase offset issue. For example, \cite{koike2009optimized} designed special PNC mapping rules for the bad relative phase offsets such that the overlapping constellation points are mapped to the same network-coded symbol. However, such specially designed PNC mapping rules only apply to the non-channel-coded case and cannot be easily extended to the channel-coded case. Certain advanced channel decoding methods proposed in \cite{wu2014analysis,wubben2010generalized,liew2015primer} can also mitigate the relative phase offset effect. However, they are generally complex iterative decoding methods that induce significant decoding latency, which is not amenable to practical implementation. Recently, machine learning approaches have been proposed to solve the network-coded symbol ambiguity problem \cite{park2021high}. By training an end-to-end TWRC using deep neural networks (DNN), a DNN-trained PNC mapping rule is used to improve the BER performance. Nevertheless, the DNN-trained PNC mapping rule is learned from a data set with only $0^\circ$ relative phase offset. Hence, as our experiments in Section \ref{sec:experiments} point out, the system performance still degrades when the actual relative phase offset differs from $0^\circ$. 


We remark that all the solutions above focus on the conventional bit-oriented communication paradigm that only aims at delivering a given information stream as quickly and reliably as possible. In contrast, this paper aims to solve the relative phase offset issue under the new semantic communication paradigm \cite{lan2021semantic}. Enabled by advanced deep learning technologies, semantic communication has recently attracted signiﬁcant attention \cite{qin2022semantic}. Various semantic communication systems were designed for the transmission of text \cite{jiang2022deep}, image \cite{bourtsoulatze2019deep}, and speech signals \cite{weng2021semantic}. In these works, the ``semantics'' of the source information is sent to the receiver, which relates to the content and meaning of the transmitted messages, rather than the exact bit stream transmission. Hence, traditional performance metrics such as BER are no longer applicable in semantic communications that evaluate system performance from a semantic level. In the context of TWRCs under bad relative phase offsets, a degraded BER performance does not mean no information is received at the receiver. This motivates us to extract the semantic meaning of transmitted messages directly rather than ensure accurate bit stream transmission.

This paper designs the first semantic communication-empowered PNC-enabled TWRC, referred to as SC-PNC TWRC. To enable the operation of PNC,  we jointly design DNN-based transceivers at the end nodes and propose a semantic PNC decoder at the relay. We use image delivery as an application example to evaluate the performance of the SC-PNC TWRC. Experimental results show that despite the bad relative phase offsets, the two nodes in an SC-PNC TWRC can still exchange accurate meaning on a semantic level. Specifically, SC-PNC can achieve a high and stable peak signal-to-noise ratio (PSNR), a metric that evaluates the reconstruction quality of images, under different signal-to-noise ratios (SNR) and relative phase offsets.

\begin{figure*}[t]
\centerline{\includegraphics[width=0.85\textwidth]{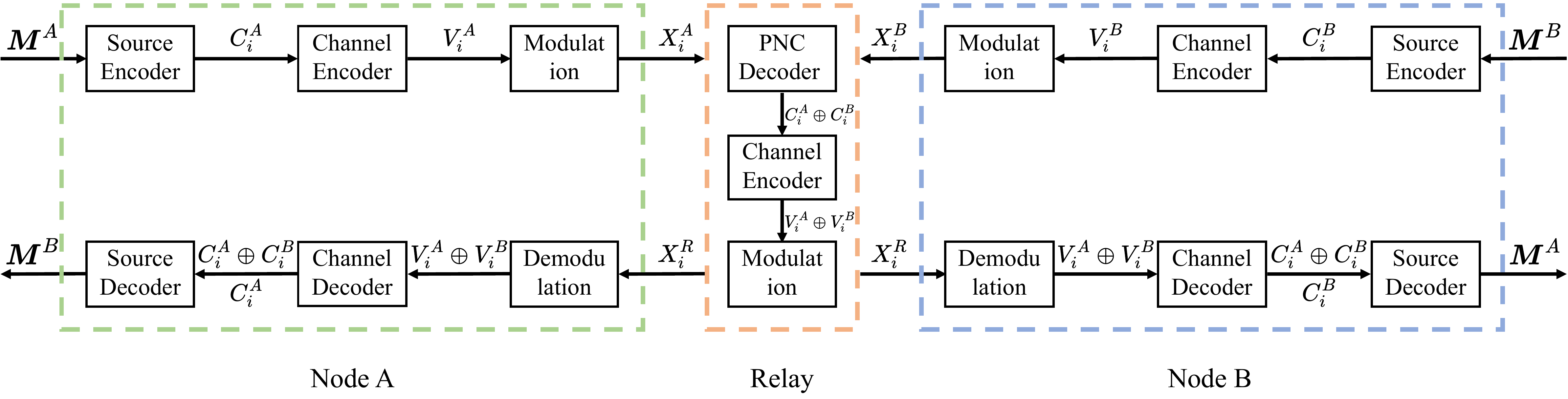}}
\vspace{-0.05in}
\caption{The general architecture of a conventional PNC-enabled TWRC.}
\label{PNC_TWRC}
\vspace{-0.2in}
\end{figure*}

\section{Conventional PNC-enabled TWRC}
\label{sec:traditional_pnc}
\subsection{System Model for the Conventional PNC-enabled TWRC}
\label{3.1}
We first review the general information processing in the conventional PNC-enabled TWRC. We assume that nodes A and B want to exchange messages $\boldsymbol{M}^{A}$ and $\boldsymbol{M}^{B}$, i.e., node A wants to send $\boldsymbol{M}^A$ to node B, and node B wants to send $\boldsymbol{M}^B$ to node A. To begin with, large messages $\boldsymbol{M}^{A}$ and $\boldsymbol{M}^{B}$ are first source-encoded into multiple packets ${C}^{A}_{i}$ and ${C}^{B}_{i}$, $i=1,2,\dots$, respectively. As an example, the messages $\boldsymbol{M}^{A}$ and $\boldsymbol{M}^{B}$ are images in this paper. In general, source encoding compresses the original message into binary bits, from which the original message can be recovered exactly. 

We consider a time-slotted system in which a packet ${C}^{A}_{i}$ or ${C}^{B}_{i}$ occupies a time slot. Next, channel coding adds redundancy bits for detecting and correcting bit errors in the transmission. Let $\Gamma(\cdot)$ denote the channel encoding operator, and we assume the use of linear channel codes in this paper. Then ${C}^{A}_{i}$ and ${C}^{B}_{i}$ are channel-encoded into ${V}^{A}_{i} = \Gamma({C}^{A}_{i})$ and ${V}^{B}_{i} = \Gamma({C}^{B}_{i})$, respectively. After channel coding, ${V}^{A}_{i}$ and ${V}^{B}_{i}$ are modulated to symbol sequences ${X}^{A}_{i}=(x^{A}_{i}[1], \dots, x^{A}_{i}[k], \dots)$ and ${X}^{B}_{i}=(x^{B}_{i}[1], \dots, x^{B}_{i}[k], \dots)$, respectively. $x^{A}_{i}\left [ k \right ]$ and $x^{B}_{i}\left [ k \right ]$ are the $k$-th modulated symbol of nodes A and B, respectively.

In an uplink slot, when nodes A and B transmit symbols ${X}^{A}_{i}$ and ${X}^{B}_{i}$ to the relay simultaneously, the received superimposed signals ${Y}^{R}_{i}=(y^{R}[1], \dots, y^{R}[k], \dots)$ at the relay can be written as
\vspace{-0.05in}
\begin{align}
{Y}^{R}_{i} = {H}^{A} \odot {X}^{A}_{i} + {H}^{B} \odot {X}^{B}_{i} + {N^R},
\end{align}
where ${H}^{A}=(h^{A}[1], \dots, h^{A}[k], \dots)$ and ${H}^{B}=(h^{B}[1], \dots, h^{B}[k], \dots)$ are the channel coefficients of the two nodes with respect to the relay, $\odot$ represents the Hadamard product, ${N^R} \sim \mathcal{CN} \left ( 0,\sigma ^2 \mathbf{I}\right ) $ indicates the independent and identically distributed (i.i.d.) complex Gaussian noise vector with variance $\sigma ^2$ at the relay, and $\mathbf{I}$ denotes a vector with the same shape as $X_{i}$ and each element is $1$. The PNC decoder attempts to decode the linear combination of ${C}^{A}_{i}$ and ${C}^{B}_{i}$, denoted by ${C}^{A}_{i}\oplus {C}^{B}_{i}$, from the superimposed signal ${Y}^{R}_{i}$, i.e., ${C}^{A}_{i}\oplus {C}^{B}_{i}$ is the eXclusive-OR (XOR) of ${C}^{A}_{i}$ and ${C}^{B}_{i}$.

In the subsequent downlink slot, the relay channel-encodes ${C}^{A}_{i}\oplus {C}^{B}_{i}$ into $\Gamma({C}^{A}_{i}\oplus {C}^{B}_{i})=\Gamma({C}^{A}_{i})\oplus\Gamma({C}^{B}_{i})={V}^{A}_{i}\oplus {V}^{B}_{i}$ (note: $\Gamma(\cdot)$ is linear). ${V}^{A}_{i}\oplus {V}^{B}_{i}$ is then modulated into ${X}^{R}_{i}$, which is broadcast to nodes A and B. The received signals at node A or B is
\vspace{-0.05in}
\begin{align}
{Y}^{J}_{i} = {H}^{J} \odot {X}^{R}_{i} + {N^J}, ~J\in\left \{A,B\right \},
\end{align}
where ${H}^{J}$ and ${N}^{J}$ represent the channel coefficient and noise term of the relay R at node $J\in\{A, B\}$, respectively. The signals received at nodes A and B are demodulated and channel-decoded into ${C}^{A}_{i}\oplus {C}^{B}_{i}$. Using ${C}^{A}_{i}\oplus {C}^{B}_{i}$, node A obtains ${C}^{B}_{i}$ by ${C}^{B}_{i} = {C}^{A}_{i}\oplus \left ( {C}^{A}_{i}\oplus {C}^{B}_{i} \right ) $. Node B follows the same manner to get ${C}^{A}_{i}$. Finally, after receiving all ${C}^{A}_{i}$ and ${C}^{B}_{i}$, $i=1,2,\dots$, original messages $\boldsymbol{M}^{A}$ and $\boldsymbol{M}^{B}$ are recovered by source decoding. As will be detailed in the next subsection, there is a relative phase offset issue that affects the overall performance of the conventional PNC-enabled TWRC. 

\subsection{The Relative Phase Offset Issue in PNC Decoding}\label{3.2}
Prior works indicated that conventional PNC decoding at the relay in the uplink phase is affected by the relative phase offset between the signals of the two simultaneously transmitted packets, especially when the modulation order is high \cite{pan2017network}. To see this, let us assume the use of QPSK modulation. We focus on the $k$-th received superimposed symbol $y^{R}_{i}[k]$, which can be expressed as
\vspace{-0.05in}
\begin{align}
y^{R}_{i}[k] = {h^A}[k]{x^{A}_{i}}[k] + {h^B}[k]{x^{B}_{i}}[k]+n[k], 
\label{eqn:superimposed_symbol}
\end{align}
where $x^{A}_{i}\left [ k \right ]$ and $x^{B}_{i}\left [ k \right ]$ are the $k$-th modulated QPSK symbols of nodes A and B, and $x^{A}_{i}\left [ k \right ], x^{B}_{i}\left [ k \right ]\in \{{1+j}, {1-j}, {-1+j}, {-1-j}\}$. An important issue in PNC decoding is how to calculate $x^{A}_{i}[k] \oplus x^B_{i}[k]$ (abbreviated as $x^{A \oplus B}_{i}[k]$) using the received sample $y^R_i[k]$ in (\ref{eqn:superimposed_symbol}). This process is referred to as \emph{PNC mapping}. According to \cite{pan2017network}, to maintain the linearity of linear channel codes $\Gamma(\cdot)$, the QPSK PNC mapping rule should be defined as
\vspace{-0.05in}
\begin{align}
x^{A \oplus B}_{i}[k] &= x^{AI}_{i}[k] \oplus x^{BI}_{i}[k] + j \cdot (x^{AQ}_{i}[k] \oplus x^{BQ}_{i}[k]),
\nonumber\\
&= x^{AI}_{i}[k]x^{BI}_{i}[k] + j \cdot (x^{AQ}_{i}[k]x^{BQ}_{i}[k]),
\label{eqn:pnc_mapping}
\end{align}
where $x^{AI}_{i}[k], x^{AQ}_{i}[k] \in \left \{-1, 1 \right \}$ ($x^{BI}_{i}[k]$, $x^{BQ}_{i}[k]$) represents the real and imaginary part of $x^{A}_{i}[k]$ ($x^{B}_{i}[k]$), respectively. 

To conveniently illustrate the effect of relative phase offset on PNC decoding, we consider a noise-free case and assume that nodes A and B have the same received power at the relay with channel gains $h^{A}[k]=1$ and $h^{B}[k]=e^{j\Delta\phi }$, where $\Delta\phi$ is the relative phase offset between the two nodes. The $k$-th superimposed symbol received at the relay is $y^{R}_{i}[k] = {x^{A}_{i}}[k] + {x^{B}_{i}}[k]e^{j\Delta\phi }$. Under some ``bad'' relative phase offsets, if the PNC mapping rule (\ref{eqn:pnc_mapping}) is used, some constellation points mapped to different network-coded symbols are inevitably overlapped with each other. For example, as pointed out in \cite{pan2017network}, when $\Delta\phi = 90^\circ$, the constellation points of symbol pairs $(1 - j, 1 - j)$ and $(1 + j, -1 - j)$ overlap at $2$, but they are mapped to $1+j$ and $-1-j$, respectively. This means that when $\Delta\phi = 90^\circ$, the QPSK mapping rule in (\ref{eqn:pnc_mapping}) can lead to symbol ambiguity, thus resulting in a degraded PNC decoding performance even in the absence of noise.

In practice, the relative phase offset is random due to uncoordinated and distributed transmitters (i.e., nodes A and B). The overall BER performance of the PNC decoder is limited by the worst-case phase offset. To deal with this issue, prior solutions focused on the conventional bit-oriented communication paradigm that cannot completely eliminate the effect of random relative phase offsets, as illustrated in Section \ref{section1}. Inspired by the new semantic communication (SC) paradigm, we design the first SC-empowered PNC-enabled TWRC in this paper, referred to as SC-PNC TWRC, to solve the relative phase offset problem. Interestingly, despite the bad relative phase offset (say, $90^\circ$), we find that the two nodes in an SC-PNC TWRC can exchange accurate meaning on a semantic level.

\section{Semantic Communication-Empowered PNC-Enabled TWRC}
\label{sec:semantic_pnc}

\subsection{Overall Framework of the SC-PNC TWRC}
As in the conventional PNC-enabled TWRC, we consider that nodes A and B exchange messages $\boldsymbol{M}^{A}$ and $\boldsymbol{M}^{B}$ (e.g., images in this paper) via relay R in the SC-PNC TWRC, whose system architecture is shown in Fig.~\ref{Semantic_PNC}. The whole system consists of three main components: $(i)$ the encoder at each node, including a semantic encoder and a channel encoder, $(ii)$ the semantic PNC decoder at relay R, and $(iii)$ the decoder at each node, including a channel decoder and a semantic decoder. We detail the three components as follows.

\subsubsection{Encoder}
In an SC-PNC TWRC, $\boldsymbol{M}^{A}$ and $\boldsymbol{M}^{B}$ are encoded into symbol sequences $X^A_{sc}$ and $X^B_{sc}$ through DNNs, which are then sent by the nodes over the wireless channel simultaneously. During the encoding procedure, two encoders are used at node A or B: the semantic encoder and the channel encoder. Specifically, the semantic encoder extracts the main features (i.e., semantic information) of the message. The channel encoder is used to encode the semantic information to mitigate wireless channel impairments and further compress the semantic information into transmit symbols. 

\begin{figure}
\centerline{\includegraphics[width=0.45\textwidth]{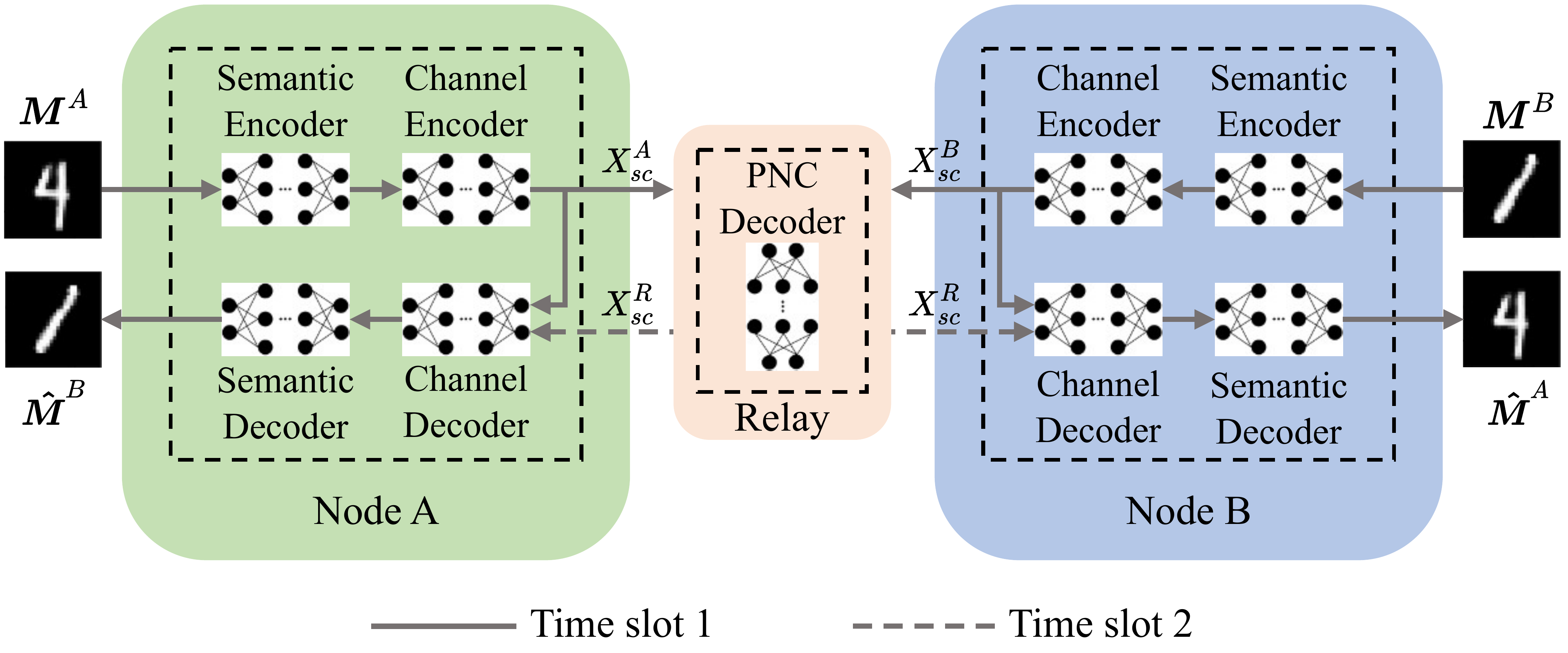}}
\vspace{-0.05in}
\caption{The overall framework of the  semantic communication-empowered PNC-enabled (SC-PNC) TWRC.}
\label{Semantic_PNC}
\vspace{-0.2in}
\end{figure}

Let us denote the training parameters of the DNNs for the semantic encoder and the channel encoder as $\alpha$ and $\beta$, respectively. Then the encoded symbol sequences $X^{J}_{sc}, J\in\left \{A,B\right \}$ can be expressed as
\vspace{-0.1in}
\begin{align}
X^{J}_{sc} = \mathbf{T}_{\beta}^{c}\left ( \mathbf{T}_{\alpha }^{s}\left ( \boldsymbol{M}^{J}\right ) \right ),
\end{align}
where $\mathbf{T}_{\alpha }^{s}\left ( \cdot \right )$ denotes the semantic encoder with parameters $\alpha$ and $\mathbf{T}_{\beta }^{c}\left ( \cdot \right )$ denotes the channel encoder with parameters $\beta$. As in the traditional PNC-enabled TWRC, nodes A and B transmit symbols $X^A_{sc}$ and $X^B_{sc}$ to the relay simultaneously in an uplink slot (see Fig.~\ref{Semantic_PNC}). Let us use  ${Y}^{R}_{sc}$ to denote the superimposed signals received by the relay in the SC-PNC TWRC. At the relay, semantic PNC decoding is performed using the superimposed signals ${Y}^{R}_{sc}$. 

\subsubsection{The semantic PNC decoder}
Unlike the conventional PNC decoder discussed in Section \ref{sec:traditional_pnc}, a key difference of the semantic PNC decoder is that the relay does not perform a PNC mapping on the received superimposed signals according to specific PNC mapping rules (such as (\ref{eqn:pnc_mapping})). Instead, as shown in Fig.~\ref{Semantic_PNC}, we adopt a DNN as the PNC decoder for mapping the superimposed signals to the network-coded symbols broadcast in the downlink. In other words, the network-coded symbols broadcast back to the end nodes are directly obtained from the output of a well-trained DNN. Denote the DNN parameters of the semantic PNC decoder as $\delta $. The symbols broadcast by the relay, $X^{R}_{sc}$, can be expressed as
\vspace{-0.05in}
\begin{align}
X^R_{sc} = \mathbf{H}_{\delta}\left ( Y^{R}_{sc}\right ),
\end{align}
where $\mathbf{H}_{\delta}\left ( \cdot \right )$ denotes the semantic PNC decoder with parameters $\delta$. As depicted in Fig.~\ref{Semantic_PNC}, nodes A and B receive signals $Y^A_{sc}$ and $Y^B_{sc}$ in the downlink, respectively.

\subsubsection{Decoder}
Similar to the encoder, the decoder also consists of two parts: the channel decoder and the semantic decoder. As in the traditional PNC-enabled TWRC, a node needs to combine the network-coded symbols received from the relay and the symbols sent by itself to recover the symbols from the other node. The channel decoder is used to realize the above procedure and obtain the semantic information of the message from the other node. Later, the semantic decoder reconstructs the message based on the semantic information. The channel decoder and semantic decoder are also implemented by two DNNs. Let $\eta $ and $\varphi $ denote their DNN parameters, respectively, so the reconstructed message is expressed as
\vspace{-0.05in}
\begin{align}
\resizebox{0.9\hsize}{!}{$
\hat{\boldsymbol{M}}^{J} = \mathbf{R}_{\varphi }^{s}\left ( \mathbf{R}_{\eta  }^{c}\left ( X^{K}_{sc},Y^{K}_{sc}\right )  \right ), \left ( J,K \right ) \in \left \{ \left ( A,B \right ),\left ( B,A \right )  \right \},$}
\end{align}
where $\mathbf{R}_{\eta }^{c}\left ( \cdot \right )$ denotes the channel decoder with parameters $\eta$ and $\mathbf{R}_{\varphi }^{s}\left ( \cdot \right )$ denotes the semantic decoder with parameters $\varphi$.

\subsection{DNN Implementation Details in the SC-PNC TWRC}

\begin{figure}[t]
\centering
\subfigure[]{\includegraphics[width=0.45\textwidth]{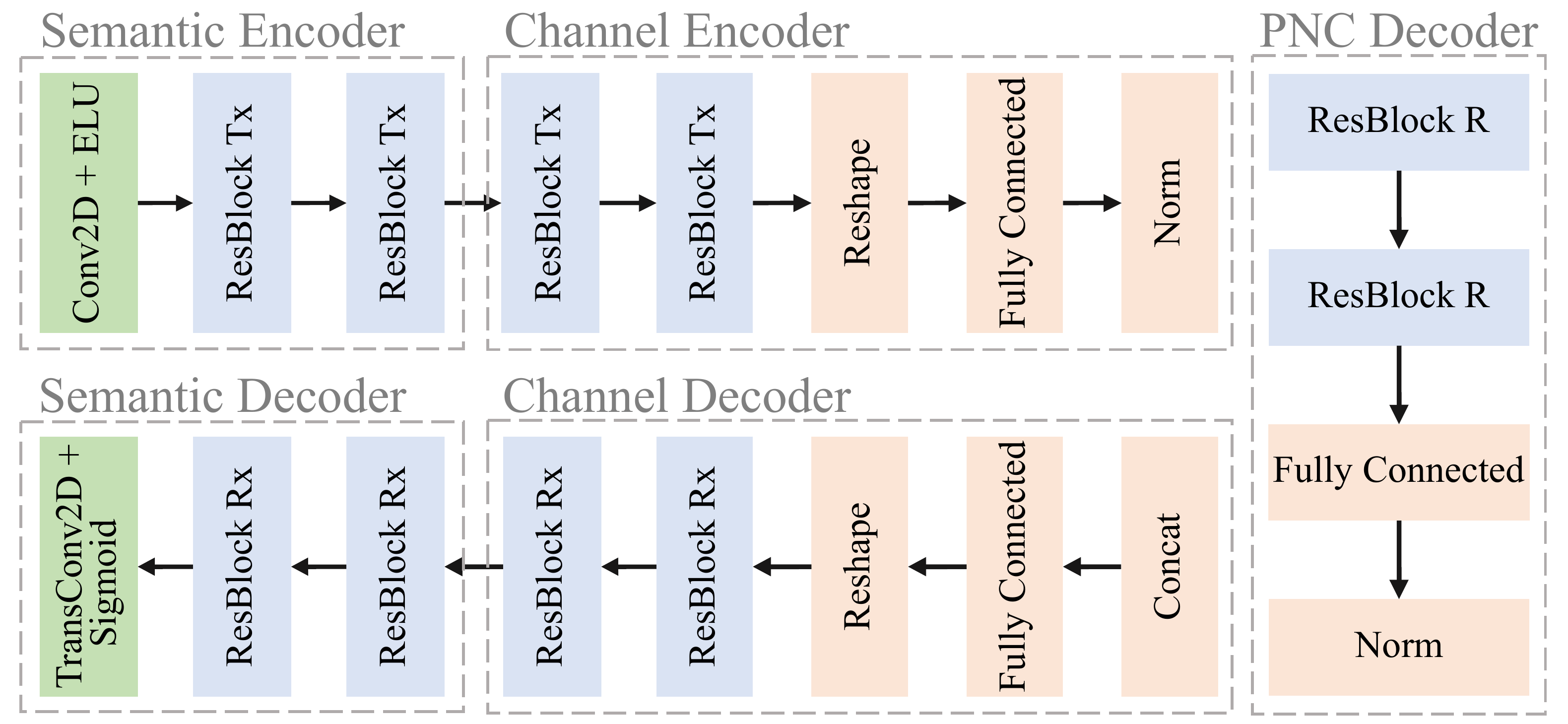}\vspace{-0.05in}\label{Detail(a)}}
\subfigure[]{\includegraphics[width=0.28\textwidth]{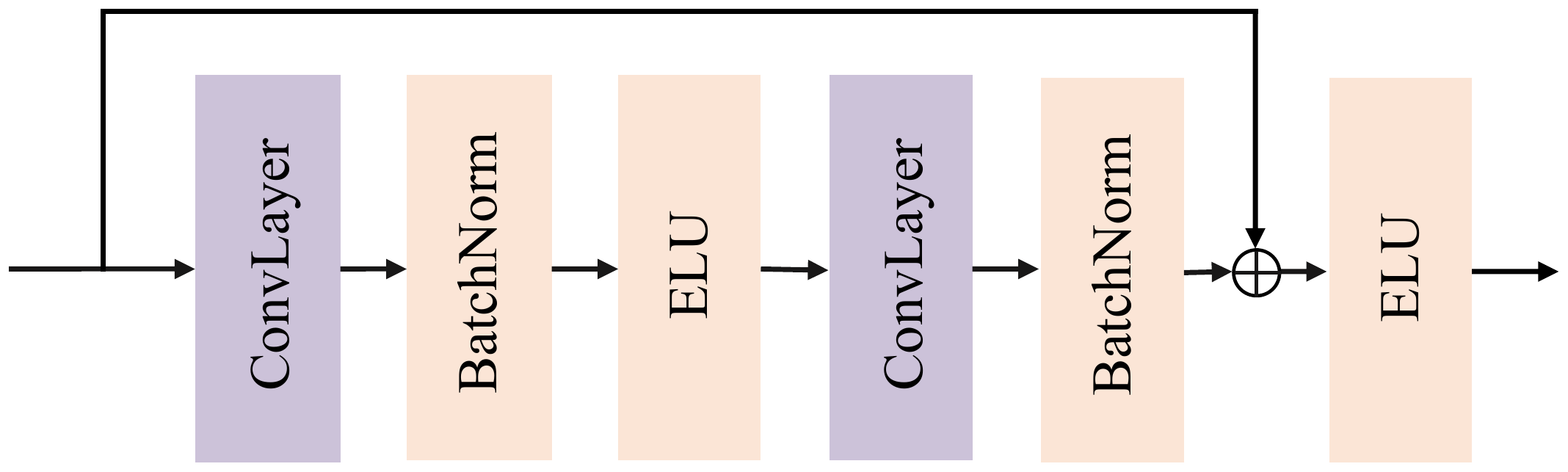}\vspace{-0.05in}\label{Detail(b)}}
\vspace{-0.1in}
\caption{(a) DNN implementation details in the SC-PNC TWRC: for ease of illustration, here we only show the information processing flow from one node to another. The two-way information processing flow can be generalized accordingly. (b) The general architecture of a residual block. }
\label{Detail}
\vspace{-0.2in}
\end{figure}

We now describe the DNN implementation of each component in the SC-PNC TWRC, as shown in Fig.~\ref{Detail}. 
First, a batch of $b$ images, $\boldsymbol{M}^J\in \mathfrak{R} ^{h \times w \times c}, J \in\{A, B\}$, are input to the encoder for training the DNNs, where $\mathfrak{R}$ denotes the set of real numbers. Here $b$ is the batch size, $h$ is the height of each image, $w$ is the width of each image, and $c$ corresponds to the number of channels of the image (i.e., $c=1$ for grayscale images, and $c=3$ for color images in RGB format). Each pixel value of the images is normalized to the range of $\left [ 0,1 \right ]$. 

As shown in Fig.~\ref{Detail(a)}, the implementation of the semantic encoder starts with a two-dimensional convolutional layer (Conv2D) with an exponential linear unit (ELU) activation function, followed by two residual blocks. Fig.~\ref{Detail(b)} shows the general architecture of a residual block. In Fig.~\ref{Detail(a)}, we denote the residual blocks used at the encoder, relay, and decoder as ``ResBlock Tx'', ``ResBlock R'', and ``ResBlock Rx'', respectively. They have slightly different  convolutional layers (ConvLayer). The ConvLayers in ResBlock Tx are Conv2D, and the detailed parameters of Conv2D in the two ResBlock Tx of the semantic encoder are summarized in Table~\ref{ConvLayer_Parameter}.
After the semantic encoder, we can obtain a batch of $b$ semantic features $f \in \mathfrak{R} ^{\frac{h}{4}  \times \frac{w}{4}  \times 16}$ of the images in our implementation.

\begin{table}[tbp]
\caption{Parameter Settings for the Semantic Enc/Decoder.}
\vspace{-0.1in}
\begin{center}
\begin{threeparttable}
\begin{tabular}{|c|c|c|c|}
\hline
\textbf{} & \textbf{ConvLayer} & \textbf{Parameter} & \textbf{Settings} \\ \hline
\multirow{3}{*}{\textbf{Semantic~Encoder}\tnote{*}} & \multirow{3}{*}{Conv2D}      & filters     & 8,8;~16,16 \\ \cline{3-4} 
 & & kernel size & 3,3;~3,3   \\ \cline{3-4} 
 & & stride      & 2,1;~2,1   \\ \hline
\multirow{3}{*}{\textbf{Semantic~Decoder}} & \multirow{3}{*}{TransConv2D} & filters     & 4,4;~8,8   \\ \cline{3-4} 
 & & kernel size & 3,3;~3,3   \\ \cline{3-4} 
 & & stride      & 1,2;~1,2   \\ \hline
\end{tabular}
\begin{tablenotes}
        \footnotesize
        \item[*] E.g., since the semantic encoder has two ResBlocks Tx, each consisting of two Conv2D, the filter parameters corresponding to the four Conv2D are 8, 8, 16, and 16, respectively.
\end{tablenotes}
\end{threeparttable}
\label{ConvLayer_Parameter}
\end{center}
\vspace{-0.2in}
\end{table}

To map the semantic feature $f$ of the images to transmitted symbols, the channel encoder also consists of two ResBlock Tx, where each ConvLayer has $32$ filters with kernels of size $3\times 3$ and the stride is size $1$. After passing through the reshape and the fully connected layers, we obtained a batch of $b$ transmitted symbols $X^{J}_{sc}\in \mathfrak{C}^{\frac{m}{2} \times 1}, J\in\{A,B\}$, by mapping the output of the fully connected layer to the real and imaginary parts of the symbols. Here $\mathfrak{C}$ denotes the set of complex numbers, and $m$ is the number of neurons in the fully connected layer. Finally, a normalization layer is required on transmitted symbols $X^{A}_{sc}(X^{B}_{sc})$ to ensure a unit transmit power constraint, i.e., $\mathbb{E} \left \| X^{J}_{sc} \right \| ^2 = 1, ~J\in\left \{A,B\right \}$.

As for the semantic PNC decoder, when the relay receives a batch of $b$ superimposed signals $Y^R_{sc} \in \mathfrak{C}^{\frac{m}{2} \times 1}$, the real and imaginary parts of the received superimposed symbols are concatenated and passed to the PNC decoder. With respect to Fig.~\ref{Detail(a)}, the semantic PNC decoder consists of two ResBlock R, whose structure is the same as ResBlock TX, except that each ConvLayer is a one-dimensional (1D) convolutional layer. Each ConvLayer in the two ResBlock R of the semantic PNC encoder has $32$ filters with kernels of size $3 \times 1$ and the stride size is $1$. 
After the fully connected and the normalization layers, the relay obtains a batch of $b$ network-coded symbols, $X^R_{sc} \in \mathfrak{C}^{\frac{m}{2} \times 1}$.

At the end nodes, the channel decoder first concatenates symbols $X^J_{sc}$ transmitted by node $J$ and the received network-coded symbols $Y^J_{sc}$ sent by the relay through a concatenate layer, which is then passed through the fully connected and the reshape layers. Later, two ResBlock Rx are used, where each ConvLayer in a ResBlock Rx is now a transposed 2D convolutional layer (TransConv2D) to restore the size of the image, and each has $16$ filters with a kernel size of $3 \times 3$ and a stride size of 1. After the channel decoder, we obtain a batch of $b$ estimated semantic features $\hat{f} \in \mathfrak{R} ^{\frac{h}{4}  \times \frac{w}{4}  \times 16}$ of the images. 

Finally, images are recovered by passing $\hat{f}$ to the semantic decoder, where the parameters of TransConv2D in the two ResBlock Rx of the semantic decoder are summarized in Table~\ref{ConvLayer_Parameter}.
Note that the activation function in the last TransConv2D of the semantic decoder is sigmoid instead of ELU in the implementation. Finally, the de-normalization layer rescales each value of the image pixels back to the range of $\left [ 0,255 \right ]$. A batch of $b$ images are reconstructed as $\hat{\boldsymbol{M}}^B(\hat{\boldsymbol{M}}^A) \in \mathfrak{R} ^{h \times w \times c}$.

\subsection{Loss Function in DNN Training}
In image delivery applications, the objective of the SC-PNC-TWRC is to reconstruct images as accurately as possible, i.e., to minimize the average distortion between a transmitted image and its reconstructed one. Hence, we adopt the mean squared error (MSE) as the loss function in the DNN training. 
Without loss of generality, we assume that the original image and the reconstructed image are grayscale images with size $h\times w$, denoted by $\boldsymbol{M}$ and $\hat{\boldsymbol{M}}$ respectively. The MSE between $\boldsymbol{M}$ and $\hat{\boldsymbol{M}}$ is
\vspace{-0.1in}
\begin{align}
MSE(\boldsymbol{M},\hat{\boldsymbol{M}}) = \frac{1}{hw}\sum^{h}_{i=1}\sum^{w}_{j=1} \left (\boldsymbol{M}_{i,j}-\hat{\boldsymbol{M}}_{i,j} \right )^2,
\end{align}
where $\boldsymbol{M}_{i,j}(\hat{\boldsymbol{M}}_{i,j})$ represents the pixel value at the position $(i,j)$ of the image. The loss function in training the SC-PNC TWRC is defined as
\vspace{-0.05in}
\begin{align}
\resizebox{0.9\hsize}{!}{$
\mathcal{L}_{MSE}=\frac{1}{N} \sum_{n=1}^{N}\left [ MSE  \left (\boldsymbol{M}^{A}_{n},\hat{\boldsymbol{M}}^{A}_{n} \right ) + MSE \left (\boldsymbol{M}^{B}_{n},\hat{\boldsymbol{M}}^{B}_{n} \right ) \right ],$}
\end{align}
where $MSE(\boldsymbol{M}^{A}_{n},\hat{\boldsymbol{M}}^{A}_{n})$  ($MSE(\boldsymbol{M}^{B}_{n},\hat{\boldsymbol{M}}^{B}_{n})$) is the MSE between the $n$-th transmitted image $\boldsymbol{M}^{A}_{n}$($\boldsymbol{M}^{B}_{n}$) and the reconstructed image $\hat{\boldsymbol{M}}^{A}_{n}$($\hat{\boldsymbol{M}}^{B}_{n}$), and $N$ is the number of images.

\section{Experimental Evaluation}
\label{sec:experiments}

\subsection{Experimental Setup}
To evaluate the SC-PNC TWRC, we consider using the MNIST handwritten digit dataset for our experiments. MNIST consists of $60,000$ training and $10,000$ testing grayscale images, each with $28 \times 28$ pixels. We implement SC-PNC TWRC using Tensorflow 2.8. The Adam optimizer~\cite{kingma2014adam} is adopted to train the DNNs. We set the training batch size to $b=128$ and fix the learning rate to $0.001$. For each training batch, each node sends $b$ MNIST training images to each other. In addition, the relative phase offset $\Delta\phi$ between nodes A and B is uniformly generated between $0^\circ$ and $360^\circ$ for each pair of $ \boldsymbol{M}_{i}^{A} $ and $ \boldsymbol{M}_{i}^{B}$. Hence, the overall training data set for SC-PNC can be represented by $\{ (\boldsymbol{M}_{i}^{A}, \boldsymbol{M}_{i}^{B}, \Delta\phi_{i} ) \}_{i=1}^{60,000}$. Additive white Gaussian channels are considered in this paper. During the training process, the SNR of each node is fixed to $7$dB in both uplink and downlink transmissions. For benchmarking purposes, we consider the following two schemes:

\textbf{Conventional PNC-enabled TWRC (Conv-PNC)}: This is the conventional PNC-enabled TWRC described in Section \ref{sec:traditional_pnc}.\footnote{For simplicity and comparison with D-PNC and SC-PNC, end nodes send the MNIST images directly without source coding. An image is simply divided into several packets, followed by channel coding and modulation.} The rate-$1/2 \left[133, 171\right]_8$ convolutional codes defined in the IEEE 802.11 standards \cite{IEEE_Standard} are used for channel coding. QPSK is used (i.e., two source bits per symbol). At the relay, the so-called XOR channel decoding (XOR-CD) \cite{pan2017network} is used to decode the PNC packet. The XOR-CD decoder maintains the linearity of convolutional codes when performing the QPSK PNC mapping (\ref{eqn:pnc_mapping}). To demonstrate the end-to-end performance of image recovery, when the relay fails to decode a PNC packet (e.g., the decoded packet does not pass the cyclic redundancy check), the relay will still forward the decoded packet in the presence of bit errors in the downlink.

\textbf{DNN-based PNC (D-PNC)}: D-PNC was proposed in \cite{park2021high}. We reproduce D-PNC using the same hyperparameter settings in \cite{park2021high}. A joint channel coding and modulation scheme is obtained by training the DNNs. Specifically, each symbol sent by the end nodes is a DNN-trained QPSK modulated symbol (instead of a regular QPSK modulated symbol in Conv-PNC), each containing two source bits. Moreover, we follow \cite{park2021high} to set the relative phase offset $\Delta\phi$ to $0^\circ$ in the training process.

Since image delivery applications are usually concerned with how similar the reconstructed image is at the receiver compared with the transmitted image, we measure the performances of different systems in terms of PSNR, which measures the squared intensity difference between the reconstructed and original image pixels. Specifically, for a grayscale image of size $h \times w$, the PSNR between the transmitted image $\boldsymbol{M}$ and the reconstructed image $\hat{\boldsymbol{M}}$ is defined as
\vspace{-0.05in}
\begin{align}
PSNR = 10 \log_{10} \left (\frac{\left (MAX_{\boldsymbol{M}}\right)^2}{MSE(\boldsymbol{M},\hat{\boldsymbol{M}})} \right)~~dB,
\end{align}
where $MAX_{\boldsymbol{M}}$ is the maximum possible pixel value of the transmitted image, and $MSE(\boldsymbol{M},\hat{\boldsymbol{M}})$ is the MSE between the transmitted image $\boldsymbol{M}$ and the reconstructed image $\hat{\boldsymbol{M}}$ (note: Conv-PNC and D-PNC directly use the received image, possibly with incorrect pixel values, as $\hat{\boldsymbol{M}}$ to calculate PSNR). If the MSE between the transmitted image and the reconstructed image is smaller, the PSNR is larger and the reconstruction quality of the image is better.

\subsection{Performance Comparison}

\begin{figure*}[t]
\centering
\subfigure[]{\includegraphics[width=0.3\textwidth]{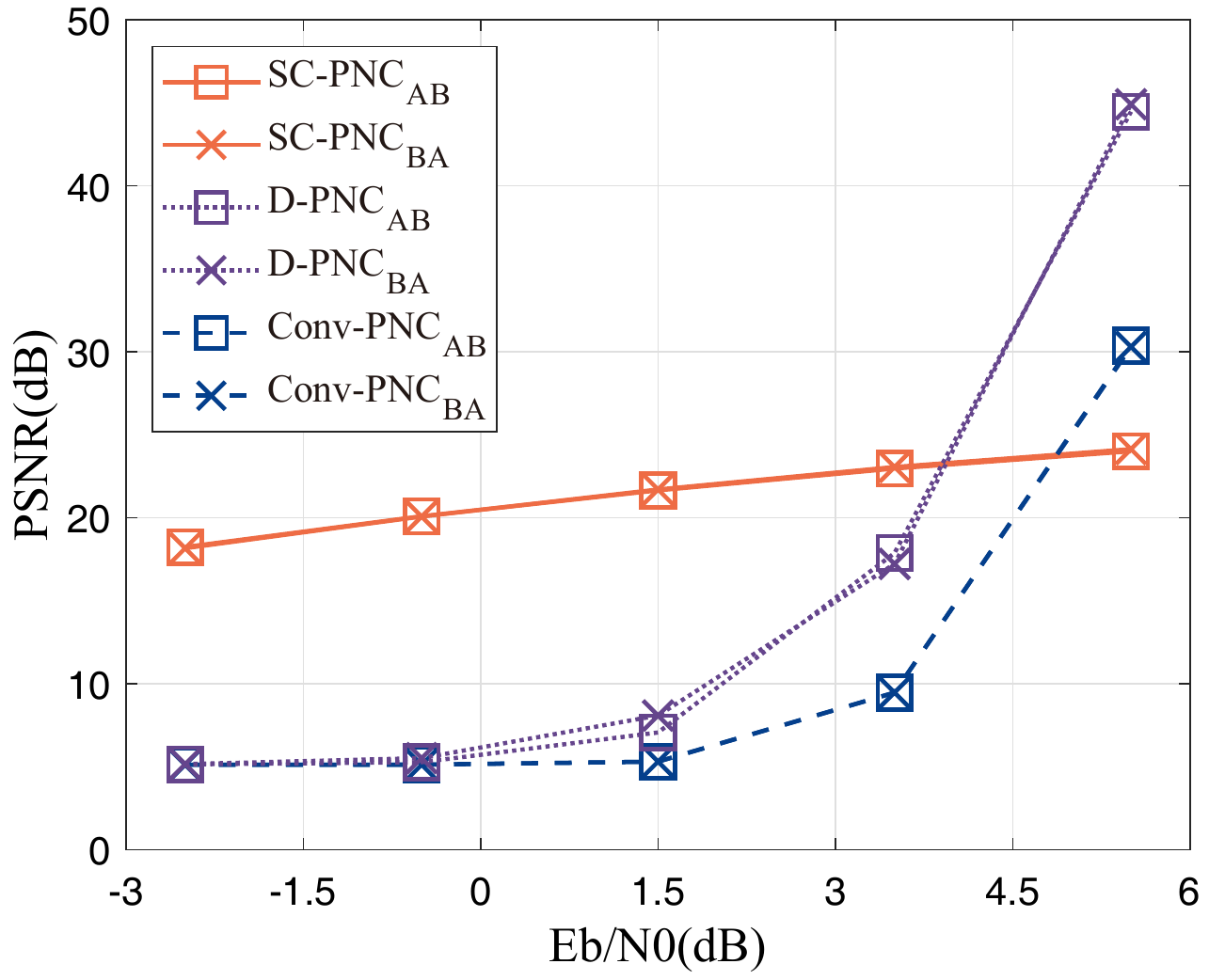}\label{snr_phi0}}
\subfigure[]{\includegraphics[width=0.3\textwidth]{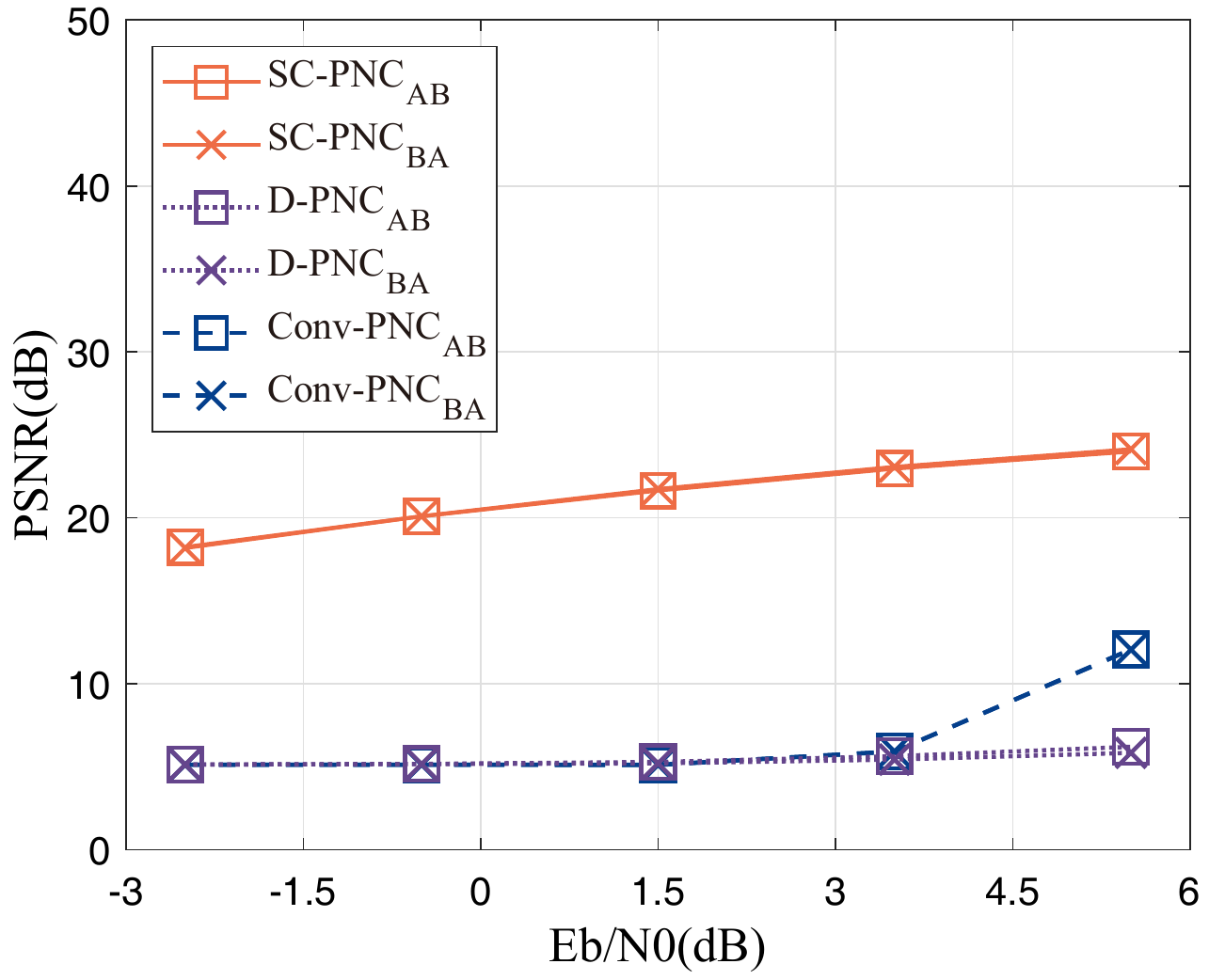}\label{snr_phi45} }
\subfigure[]{\includegraphics[width=0.3\textwidth]{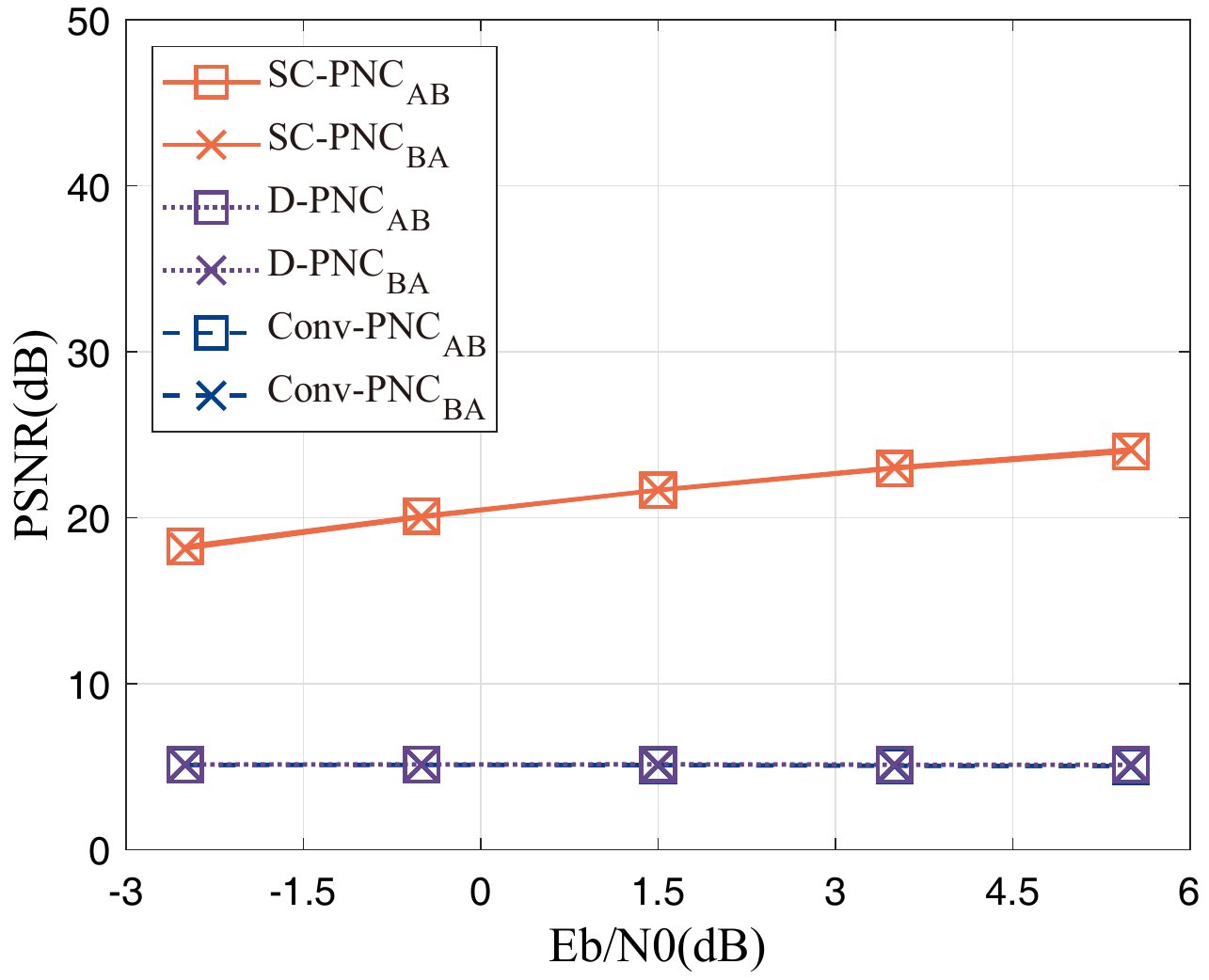}\label{snr_phi90}}
\vspace{-0.1in}
\caption{PSNR performance of Conv-PNC, D-PNC, and our proposed SC-PNC under different SNRs when the relative phase offsets are (a) $0^\circ$, (b) $45^\circ$, and (c) $90^\circ$. }
\label{SNR_PSNR}
\vspace{-0.2in}
\end{figure*}


Fig.~\ref{SNR_PSNR} plots the PSNR of Conv-PNC, D-PNC, and SC-PNC under different SNRs, when the relative phase offset $\Delta\phi$ is (a) $0^\circ$, (b) $45^\circ$, and (c) $90^\circ$.  We consider the PSNR performance of the two end nodes separately. When testing the performance of the three systems, we consider an SNR-balanced scenario in which nodes A and B have the same SNR in both uplink and downlink transmissions. We vary the SNR from $-3$dB to $6$dB. Notice that the SNRs in testing are different from the SNR in training (i.e., $7$dB). For the legends in Fig.~\ref{SNR_PSNR}, as an example, SC-PNC$\mathrm{_{AB}}$ represents SC-PNC when node A sends to node B, i.e., the PSNR is measured at node B. 

Fig.~\ref{SNR_PSNR} shows that the PSNR increases with increasing SNR for all three systems. In particular, when the relative phase offset $\Delta\phi$ is $0^\circ$ (see Fig.~\ref{snr_phi0}), D-PNC outperforms Conv-PNC because D-PNC jointly trains the channel coding and modulation scheme by deep learning and achieves lower BERs than the separately designed schemes in Conv-PNC. Furthermore, SC-PNC outperforms D-PNC when the SNR is low. At low SNRs, D-PNC suffers from high BERs due to poor channel conditions, resulting in low PSNRs. However, SC-PNC can still recover useful semantic information at low SNRs. In addition, D-PNC performs better than SC-PNC when the SNR is high, because D-PNC can reconstruct the transmitted image with accurate pixel values (i.e., without bit errors). By contrast, SC-PNC recovers images at the semantic level and does not obtain exact pixel values of the original image as D-PNC does. This phenomenon is consistent with previous works on semantic communication~\cite{bourtsoulatze2019deep}.

Fig.~\ref{snr_phi45} and Fig.~\ref{snr_phi90} show the PSNR performance against SNR when the relative phase offsets $\Delta\phi$ are $45^\circ$ and $90^\circ$. As shown in both figures, Conv-PNC and D-PNC suffer from performance degradation compared with Fig.~\ref{snr_phi0} when $\Delta\phi$ is $0^\circ$. For Conv-PNC, the increase in $\Delta\phi$ reduces the Euclidean distance between constellation points mapped to different XOR symbols, thus degrading the BER performance \cite{pan2017network} (lowering the PSNR as well). For example, when $\Delta\phi$ is $90^\circ$, there are different XOR symbols overlapping each other with a Euclidean distance of zero. This leads to XOR symbol ambiguity when the QPSK PNC mapping rule (\ref{eqn:pnc_mapping}) is used. For D-PNC, it only considers $\Delta\phi=0^\circ$ during the training process. When the actual $\Delta\phi$ is different from the one in training, the joint channel coding and modulation scheme learned from the training data does not work well. In other words, to achieve a good PSNR performance, we need to re-train the DNNs of D-PNC to facilitate different relative phase offsets. 

We see from Fig.~\ref{snr_phi45} and Fig.~\ref{snr_phi90} that SC-PNC outperforms both Conv-PNC and D-PNC under all SNR values. Compared with Conv-PNC (with the QPSK PNC mapping (\ref{eqn:pnc_mapping})), which leads to XOR symbol ambiguity when $\Delta\phi$ is $90^\circ$, SC-PNC does not rely on a specific PNC mapping rule and learns an appropriate rule via DNNs, thus avoiding the symbol ambiguity problem and improving the PSNR performance. While D-PNC also learns the PNC mapping via DNNs, it is designed based on minimizing BERs. When the trained $\Delta\phi$ is different from the tested $\Delta\phi$, an improper PNC mapping learned from DNNs leads to a degraded BER performance. With the new semantic communication paradigm, SC-PNC introduces the semantic encoder/decoder jointly trained with the channel encoder/decoder in the transceiver design. By doing so, messages can be directly recovered at the semantic level instead of the bit level. At the semantic level, our experimental results show that there is no need to worry about the disparity between the trained and the tested relative phase offsets. For example, when we generate $\Delta\phi$ randomly between $0^\circ$ and $360^\circ$ in the training process, Fig.~\ref{SNR_PSNR} indicates that SC-PNC can achieve high and stable PSNR performance under different SNRs and relative phase offsets.\footnote{To be fair, we also try to train D-PNC by generating $\Delta\phi$ uniformly between $0^\circ$ and $360^\circ$ for each pair of $ \boldsymbol{M}_{i}^{A} $ and $ \boldsymbol{M}_{i}^{B}$ as SC-PNC does. However, this approach leads to poor BER/PSNR performances under all $\Delta\phi$. This indicates that a unified joint channel coding and modulation scheme aimed at low BERs cannot be learned by DNNs when the relative phase offset varies. By contrast, SC-PNC performs well because it recovers the semantic meaning of the images rather than the exact bit information.}

\section{Conclusion}
\label{sec:conclusion}
We have presented SC-PNC TWRC, the first semantic communication-empowered PNC-enabled TWRC. In particular, we have demonstrated the feasibility of using the new semantic communication paradigm to tackle the bad relative phase offset problem in PNC decoding. With the help of DNNs, our designed SC-PNC TWRC realizes semantic PNC decoding at the relay and the direct extraction of semantic information at the end nodes. This solves the performance degradation problem in the conventional bit-oriented communication design that only aims to deliver bit streams reliably. Experimental results show that the SC-PNC TWRC outperforms its bit-oriented benchmarks by achieving high and stable PSNR performance at different relative phase offsets in image delivery applications.




\end{document}